# Controlling the Bandwidth of High Harmonic Emission Peaks with the Spectral Polarization of the Driver

ELDAR RAGONIS[1,2], ERAN BEN-AROSH[1,2], LEV MERENSKY[1,2] AND AVNER FLEISCHER[1,2,*]

[1]*Raymond and Beverly Sackler Faculty of Exact Sciences, School of Chemistry, Tel Aviv University,Tel Aviv 6997801, Israel*
[2]*Tel-Aviv University center for Light-Matter-Interaction, Tel Aviv 6997801, Israel*
*Corresponding author: avnerfleisch@tauex.tau.ac.il

**We demonstrate a High-Harmonic-Generation scheme which offers control over the bandwidth of the spectral peaks. The scheme uses a vectorial two-color driver with close central frequencies, generated by spectrally splitting a linearly-polarized input femtosecond-duration laser pulse and subsequent recombining the two halves after their polarizations are made cross-elliptical and counter-rotating. This results in the generation of new emission channels which coalesce into broad odd-integer HHG peaks, the bandwidth of each being proportional to the frequency difference between the two colors, to the harmonic order and inversely-proportional to the driver fields' ellipticities. Peak-broadening to the extent that a supercontinuum is formed is also demonstrated. This source will find use in HHG applications benefiting from high-flux broadband extreme ultra-violet radiation, such as attosecond transient absorption spectroscopy.**

High Harmonic Generation (HHG) offers a key tabletop source of extreme ultra violet (XUV) radiation for studies ranging from attosecond science and time-resolved exploration of ultrafast electron dynamics in matter [1-3] to imaging of nanostructures [4-6]. Driving the process by a quasi-monochromatic driver results in the emission of a sparse comb of odd-integer multiples of the driver frequency. Many applications however, e.g. attosecond transient absorption spectroscopy (ATAS) [7-9], require XUV radiation at wavelengths which do not necessarily appear in the comb. In order to increase and control the bandwidth of the peaks in the HHG spectra (HGS), the number of induced electron recollisions needs to be controlled. This could be done by shortening the driver's duration [10] by post-compression techniques ([11]) which however work best with femtosecond input beams with average power of few Watts ([12]). With very high or very low average-power laser sources, polarization-gating [13-18] or time-gating using a two-color scheme [19-20] could be used. On the other hand, the pulse-shortening approach offers simple control over the peak bandwidth (by dispersion management), while this is not so for the gating approach, where it requires the replacement of birefringent crystals. Simple peak bandwidth control schemes are hence needed in cases when the above approaches might impose limitations.

Here we add the polarization (spin angular momentum, SAM [21]) degree of freedom to the time-gating technique to obtain a mixed time-polarization-gating technique, which offers continuous controllability over the bandwidth of the harmonic peaks. In our scheme HHG is driven by a vectorial two-color driver with close central frequencies [22], generated by spectrally splitting a linearly-polarized input femtosecond-duration pulse from a commercial amplified Ti:Sa laser and subsequent recombining the two halves after their polarizations are made cross-elliptical and counter-rotating. This considerably broadens the odd-integer peaks in the HGS. The results are easily explained using a frequency-domain description using photonic channels. The equivalent, but less intuitive, time-domain description (utilizing the two-dimensional electron trajectories generated by the vectorial field) is described in [23].

A Gaussian pulse whose spectrum $S(\omega) = exp[-a(\omega - \omega_0)^2]$ is sufficiently narrow (i.e., $a\omega_0^2 \gg 1$) has the central frequency $\langle\omega\rangle = \int_0^\infty d\omega\, \omega S(\omega) \Big/ \int_0^\infty d\omega\, S(\omega) = \omega_0$. When such linearly-polarized input pulse is used to drive HHG, harmonics which are odd-integer multiples of the driver's central frequency are generated $\Omega = (2n - 1)\omega_0$ ($n$ is an integer). The spectral width of each harmonic peak is inversely-proportional to the number of electron recollisions occurring in the process, which is determined by the driver's spectral width. Suppose now this pulse enters a Mach-Zehnder interferometer where the common spectrally-flat beamsplitters are replaced by hard-edge dichroic mirrors which spectrally split (and later recombine) the pulse at $\omega_0$. Then two pulses exit the interferometer, the central frequency of each being

$$\omega_1 \equiv \int_{\omega_0}^\infty d\omega\, \omega S(\omega) \Big/ \int_{\omega_0}^\infty d\omega\, S(\omega) = \omega_0 + \sqrt{\frac{1}{\pi a}}$$

and

$$\omega_2 \equiv \int_0^{\omega_0} d\omega\, \omega S(\omega) \Big/ \int_0^{\omega_0} d\omega\, S(\omega) = \omega_0 - \sqrt{\frac{1}{\pi a}}.$$

When a non-zero time delay is chosen and the two pulses are overlapped in time, a two-color field is obtained (rather than the original input pulse). When this field is used to drive HHG, selection rules dictate that the following harmonic emission channels are allowed [19,21]:

$$\Omega_{(n_1, n_2)} = n_1\omega_1 + n_2\omega_2 \quad , \quad n_1 + n_2 = 2n - 1$$
$$= (n_1 + n_2)\omega_0 + (n_1 - n_2)\sqrt{1/\pi a} \quad (1)$$

where $n_1, n_2$ are integers which refer to the number of annihilated photons from each color $\omega_1$, $\omega_2$ composing the driver. Hence, this simple procedure of spectrally splitting-and-recombining (after

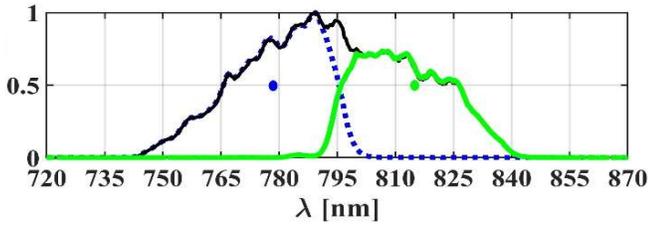

Fig. 1. Spectrum of the Ti:Sa laser input pulses entering (black) and exiting (arm 1- dashed blue, arm 2- solid green) the interferometer with the blue and green dots denoting the central wavelength $\lambda_1$, $\lambda_2$

slight temporal mismatch) a monochromatic input pulse replaces in the HGS every odd-integer $(2n-1)^{th}$ harmonic that would have been obtained were the HHG process driven directly by the input pulse, with the following set of $2n$ channels (close peaks) $\Omega_{(0,2n-1)}, \ldots, \Omega_{(n,n-1)}, \Omega_{(n+1,n-2)}, \Omega_{(n+2,n-3)}, \ldots, \Omega_{(2n-1,0)}$ when driven by the bichromatic driver. The spacing between each two channels $2\sqrt{1/\pi a}$ is determined by the value of $a$. For instance, when $a$ is large (small bandwidth of the input pulse), the two frequencies $\omega_1, \omega_2$ are close, and the spacing is small. Due to the finite duration of the driver pulses, each of these channels has some finite spectral width $A_{2n-1}$ and since the two frequencies $\omega_1, \omega_2$ are close, all channels are partially-overlapping in the HGS, such that they merge, for any practical purpose, into a single, broad peak around the $(2n-1)^{th}$ harmonic. Each broad odd-integer peak hence covers a bandwidth of $B_{2n-1} = A_{2n-1} + 2(2n-1)\sqrt{1/\pi a}$ which, according to Eq.1, increases with harmonic order (large n), the frequency difference $\omega_1 - \omega_2$ or the bandwidth of the input pulse (with large bandwidth corresponding to small $a$).

To demonstrate this experimentally, a commercial amplified Ti:Sa laser (Legend-Elite Duo HE+ USX-5kHz by *Coherent*) which delivers linearly-polarized (horizontal polarization) 25fs, 3mJ pulses at repetition rate of 5kHz was used. The laser input pulses were spectrally split and recombined in a Mach-Zehnder interferometer whose spectrally-flat beamsplitters were replaced with hard-edge short-pass dichroic mirrors with cutoff wavelength at 795nm, around the central wavelength of the Ti:Sa pulses ( $\lambda_0 = 2\pi c/\omega_0 = 796.1 nm$, $\omega_0 = 1.557 eV$ ). The interferometer generated two pulses out of each incoming input pulse, with central wavelengths of $\lambda_1 = 777.8 nm$ and $\lambda_2 = 815.2 nm$, respectively yielding a detuning of $\sqrt{1/\pi a} \approx 0.0235 \omega_0$. The spectra of the input and two exit pulses are shown in Fig.1. At the exit of the interferometer the beams were spatially and temporally overlapped (arm 1 is placed on a linear stage and the time delay was set not at zero, but where maximal HHG signal was obtained), forming a two-color driver. However, in order to add controllability to this bichromatic HHG scheme, the polarization degree of freedom [21] has been utilized, by adding two optical elements. First, a super-achromatic half-waveplate (*B. Halle*) was placed in arm 1 of the interferometer, and flipped the linear polarization from horizontal to vertical. Second, at the exit of the interferometer, prior to the HHG generation chamber, the combined beam passed through a super-achromatic quarter-waveplate (*B. Halle*) placed on a motorized rotation stage. As we will show, the orientation of this waveplate $\alpha$ controls the HGS obtained, as the possible emission channels are dictated by SAM conservation law.

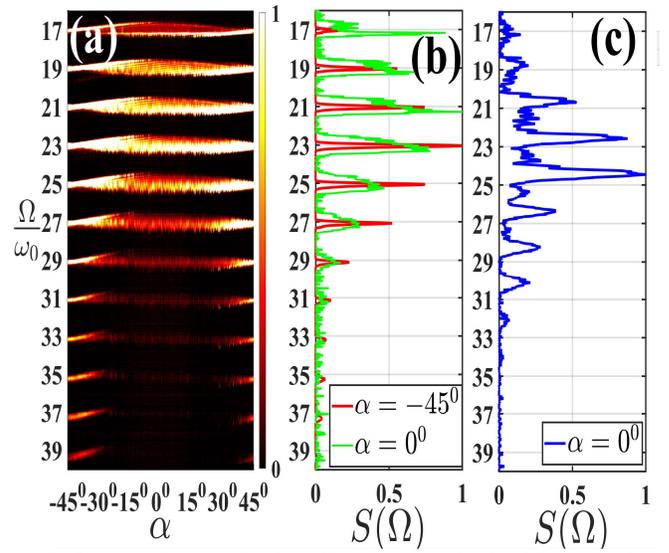

Fig. 2. (linear scale, normalized to 60000counts/sec) (a) HGS intensity S(Ω) vs. harmonic order and angle $\alpha$ of the achromatic quarter waveplate; The broad Cooper minima in Argon, around the 33rd harmonic is also visible. (b) HGS lineouts from (a) at $\alpha = -45^0$ (red curve) and $\alpha = 0^0$ (green).. (c) HGS lineout for $\alpha = 0^0$ when broadband input pulse is used (see text).

Using Second Harmonic-Generation Frequency-Resolved Optical Gating (SHG-FROG) the pulse durations from each arm just prior to the HHG chamber entrance window were measured to be 45fs and 55fs respectively. The energy per pulse in each arm was 440μJ and 480μJ respectively. The two- color driver was focused (using a f=750mm focusing mirror), few millimeters before a 100μm Argon jet in order to favor the HHG emission from the short trajectories, yielding estimated intensities [derived from the cutoff location of the HGS, around the 39th harmonic] of $1.3 \cdot 10^{14} \left[ W/cm^2 \right]$ and $1.1 \cdot 10^{14} \left[ W/cm^2 \right]$. The resulting XUV radiation was filtered from the remaining infra-red driver radiation with a 0.2μm-thick Aluminum foil. It then entered a home-built HHG spectrometer comprised of a 50μm-by-6mm vertical entrance slit (located 1365mm after the focus), aberration-corrected flat-field concave grating with 300 lines/mm (30-006 by *Shimadzu*), and a back-illuminated CCD camera (Newton DO940P-BN by *Andor*). The spectrometer had a typical resolution of 25meV at 30eV and 90meV at 70eV. When driven directly by 25fs linearly-polarized input pulses., peak bandwidth of $A_{2n-1=19} = 0.067 eV \approx 0.047 \omega_0$ was obtained for the 19th harmonic, which is about twice the detuning.. This is also the expected peak width of every harmonic channel formed in our two-color scheme in the vicinity of the 19th harmonic.

Fig. 2a shows the obtained HGS as function of harmonic order and the quarter-waveplate reading $\alpha$. For $\alpha = -45^0$ the driver is bicircular and counter-rotating (the two colors are circularly-polarized with opposite helicities). SAM conservation considerations dictate that only two harmonic channels are possible in this setting: $\Omega_{(n-1,n)}, \Omega_{(n,n-1)}$, being circularly-polarized with opposite helicities. As the reading $\alpha$ increases (e.g., $\alpha = -30^0$ corresponds to bielliptical counter rotating driver, with orthogonal major axes of the polarization ellipses), each color becomes elliptically-polarized, and supplies the system with circularly-polarized photons of two possible opposite helicities. In the basis of circular polarization, each channel is then characterized by four integers $\left(n_1^+, n_1^-, n_2^+, n_2^-\right)$ which refer to the number of annihilated photons from each arm (subscript) and each helicity (superscript). The energy and spin state of harmonic channel $\left(n_1^+, n_1^-, n_2^+, n_2^-\right)$ are respectively

$$\Omega_{\left(n_1^+, n_1^-, n_2^+, n_2^-\right)} = \left(n_1^+ + n_1^-\right)\omega_1 + \left(n_2^+ + n_2^-\right)\omega_2$$

and

$$\sigma_{\left(n_1^+, n_1^-, n_2^+, n_2^-\right)} = \left(n_1^+ + n_2^+\right) - \left(n_1^- + n_2^-\right).$$

Hence, the two channels obtained for $\alpha = -45^0$ are e.g. $\Omega_{(n-1,0,0,n)}, \Omega_{(n,0,0,n-1)}$ [and for $\alpha = 45^0$ are $\Omega_{(0,n-1,n,0)}, \Omega_{(0,n,n-1,0)}$]. These two channels are merged into a single peak (see Fig. 1 in [22]) whose measured (anticipated) bandwidth, at the 19th harmonic, is 0.20eV (0.14eV). As $\alpha$ progresses from $\alpha = -45^0$ towards $\alpha = 0^0$, higher- and lower-energy channels are added. At $\alpha = 0^0$ (cross-orthogonal linear-polarizations of the two colors) the maximal number of channels is allowed, extending all the way from $\Omega_{(0,0,n-1,n)}, \Omega_{(0,0,n,n-1)}$ to $\Omega_{(n,n-1,0,0)}, \Omega_{(n-1,n,0,0)}$ and the emission support $B_{2n-1}\left(\alpha = 0^0\right)$ is the largest. This is evident also from Fig.2b in which lineouts from Fig. 2a are shown. With average detuning of $0.0235\omega_0$ the anticipated bandwidth covered by the set of 2n channels is $B_{2n-1}\left(\alpha = 0^0\right) = A_{2n-1} + (2n-1)0.047\omega_0 \approx 2n \cdot 0.047\omega_0$, about 2n-times larger than that of each channel alone $A_{2n-1}$. The measured (anticipated) maximal bandwidth of the 19th harmonic, is however about 1.18eV (1.46eV) [same as one obtained when HHG is driven directly by linearly-polarized 7fs input pulses]. This indicates that not all channels appear in the HGS. This is not surprising, due to the importance of propensity rules (on top of selection rules) in determining the appearance likelihood of a channel. Due to photon statistics, the appearance of channels which best utilize the available photons supplied by the drivers is favored. For instance channels requiring more photons from the more intense arm on one hand, or having a small excess of photons from the two frequencies on the other hand, will be favored. For this reason, the channels constituting the edges of the peak, namely $\Omega_{(0,0,n-1,n)}, \Omega_{(0,0,n,n-1)}$, $\Omega_{(n,n-1,0,0)}, \Omega_{(n-1,n,0,0)}$ are prone to be absent, since they always utilize photons of a single color only.

Repeating the scans for different delays of the moving arm of the interferometer didn't change the results significantly, in accordance with theoretical analysis of the vectorial driver [23], which show that the relative phase between the two colors of the driver merely causes a uniform shifting of all recollision events in time, but has no effect on their relative orientations or timings. This manifests itself at most in the spectral phase of the HHG radiation but not in the HGS.

According to Eq.1, further Increase of the HHG peak bandwidth requires a larger frequency difference $\omega_1 - \omega_2$ between the pulses exiting the interferometer. This could be achieved by using a pulse-shaper, or by designing an asymmetric interferometer, in which two different dichroic mirrors are used, with cutoff frequencies $\omega_{c1}, \omega_{c2}$. This would yield the following frequencies for the arms:

$$\omega_1 = \omega_0 + \sqrt{\frac{1}{\pi a}} \frac{\exp\left(-x_1^2\right)}{1-erf(x_1)} \text{ and } \omega_2 = \omega_0 - \sqrt{\frac{1}{\pi a}} \frac{\exp\left(-x_2^2\right)}{1+erf(x_2)}$$

where $x_{1,2} = \sqrt{a}\left(\omega_{c1,2} - \omega_0\right)$ and $erf$ is the error function. For instance, dichroic mirrors with cutoff wavelengths of 779.6nm and 813.4nm would yield central wavelengths of 766.6nm, 827.9nm for the two arms, which would cause the overlapping of the high-energy and low-energy edges of the broad peaks belonging to the 25th and 27th harmonics respectively. Hence the broad peaks of all odd-harmonics above the 25th harmonic will themselves coalesce into a single giant supercontinuous peak in the HGS. According to Eq. 1, similar effect can be achieved by injecting input pulses of larger bandwidth (smaller $a$) into our existing (symmetric) interferometer. We have verified this by generating short, 11fs pulses in a Neon-filled, 500μm-diameter, 1-meter-long hollow-core capillary (HCF) [24] and subsequent compression on chirped mirrors (PC70 by *Ultrafast Innovations*). The two driver pulses exiting the interferometer had central wavelengths of $\lambda_1 = 752nm$ and $\lambda_2 = 825nm$ (yielding a detuning of $0.0463\omega_0$), same energy (150μJ) and pulse durations of 16fs and 28fs respectively. As is evident from Fig.2c, Fig. 3a this indeed extended the span of harmonic channels such that channels emerging from adjacent odd-integer harmonics got nested one into another and a supercontinuum was obtained between the 18th and 27th harmonics for $\alpha = 0^0$. For comparison, driving HHG directly with the short linearly-polarized 11fs driver pulses yielded odd-integer peaks only, with bandwidth of only 0.6eV for each peak.

A zoom-in view into the 19th – 23rd harmonics is shown in Fig.3a,. As $\alpha$ is varied, each group of channels emerging from the 19-[(2n-1)-] th harmonic forms a structure of 10 *(n)* upward-tilted bright lines crossing with 10 *(n)* downward-tilted bright lines, forming a chessboard pattern. Indeed, 10, 11 and 12 lines are easily seen for harmonics 19, 21 and 23, respectively. The diagram in Fig. 3b shows the appearance of harmonic channels (around the 7th harmonic, for simplicity) where the vertical axis represents the energy of the channel (in units of the detuning) and the horizontal axis is the reading $\alpha$ and is in proportion to the number of new circular driver photons $n_1^-, n_2^+$. The appearance of the chessboard pattern results from merging (see [23] and Fig. 1 in [22]) of every two harmonic channels [red dots, with the values of $\left(n_1^+, n_1^-, n_2^+, n_2^-\right)$ specified above or below the dot] with same values of $n_1^-, n_2^+$ or $n_1^+, n_2^-$ into a merged peak (orange dots). These merged peaks are the bright cells comprising the chessboard pattern. Due to limited dynamic range of the spectrometer harmonic channels with small appearance probability might be absent from the HGS [e.g., channels $(4,3,0,0), (0,0,3,4)$ in the diagram or channels $(10,9,0,0), (0,0,9,10)$ from the HGS in Fig. 2a]. Also, the appearance of the structure of upward/downward lines (orange strips in Fig. 3b) depends on the relative intensities of the merged peaks. For this reason the structure of lines is usually most easily observed for the lowest harmonics. The measured energy separation of every two parallel bright lines in Fig. 3a is $0.200\omega_0$, in agreement with $4 \cdot 0.0463\omega_0 \approx 0.185\omega_0$ (as anticipated by our theory). Hence, the frequency-domain photonic-channel analysis correctly describes the experimental findings.

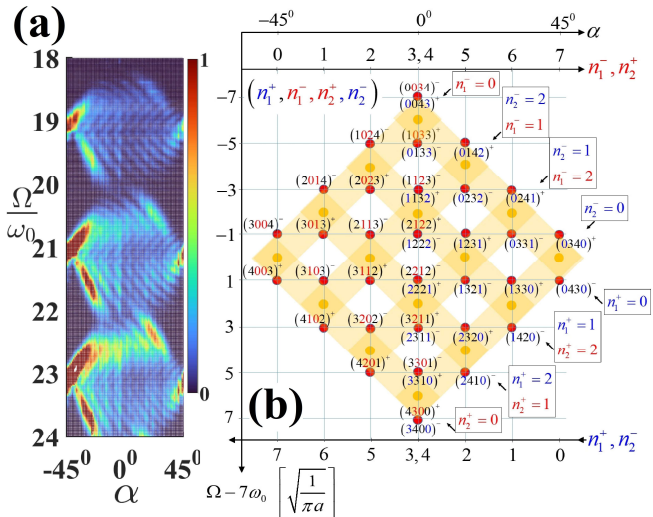

Fig. 3. (a) zoom-in view on the 19$^{th}$, 21$^{st}$ and 23$^{rd}$ harmonics obtained by our two-color scheme with short input pulses. (linear scale, normalized to 60000counts/sec) (b) Diagram representing the birth of harmonic channels (red dots) around the 7$^{th}$ harmonic, and their mergence (orange dots) into lines (orange strips). The upper horizontal axes count the number of new type of photons $n_1^-, n_2^+$ supplied to the HHG process as $\alpha$ varies.

Finally, we exclude alternative explanations for the observed peak broadening. The moderate laser intensities used, together with absence of any noticeable HGS blueshift, redshift or peak broadening [25-28] of the channels (even for the cases $\alpha = \pm 45^0$) suggest that the intensity-dependent atomic phase is not the cause for the broadening, and that spatio-spectral effects are negligible [29]. Moreover, a synthesized Fourier-transform-limited vectorial driver (corresponding to the spectra in Fig. 1 assuming flat spectral phase) yields roughly the same peak intensity [for those instants relevant for recollision (where the field polarization is close to linear)] for all values of $\alpha$. This excludes maker fringes [30] as a possible cause for the peak broadening. The focusing geometry favors the generation of the short quantum trajectories, which rules out broadening due to contribution of the long trajectories [31] or quantum path interferences [32]. Hence, the observed broadening reflects a genuine generation mechanism of new photonic channels.

To conclude, using a cross-elliptical, counter-rotating, two-color driver we have presented a method to control the bandwidth of odd-integer harmonic peaks in the HGS. Dictated by spin conservation, the number of allowed harmonic channels is varied by a simple rotation of a waveplate. These channels merge into broad peaks. Our method has been analyzed using frequency-domain considerations. In the equivalent time-domain picture [23] it is a mixed time-polarization-gating technique.

**Funding.** Israel Science Foundation (524/19).

**Acknowledgments.** We thank Prof. Oren Cohen and Dr. Pavel Sidorenko for assistance with the FROG Ptychographic retrieval algorithm (https://oren.net.technion.ac.il/ptych-frog/).

**Disclosures.** The authors declare no conflicts of interest.

**Data availability.** Data underlying the results presented in this paper may be obtained from the authors upon reasonable request.

## References

1. P. Corkum and F. Krausz, Nat. Phys. **3** 381 (2007).
2. F. Krausz and M. Ivanov, Rev. Mod. Phys. **81** 545 (2009).
3. J. Li, J. Lu, A. Chew, S. Han, J. Li, Y. Wu, H. Wang, S. Ghimire and Z. Chang, Nat. Comm. 11 2748 (2020).
4. O. Kfir, S. Zayko, C. Nolte, M. Sivis, M. Möller, B. Hebler, S. S. P. K. Arekapudi, D. Steil, S. Schäfer, M. Albrecht, O. Cohen, S. Mathias and Claus Ropers, Science advances 3 eaao4641 (2017).
5. J. Miao, T. Ishikawa, I. K. Robinson, M. M. Murnane, Science 348, 530 (2015).
6. E. R. Shanblatt, C. L. Porter, D. F. Gardner, G. F. Mancini, R. M. Karl, Jr., M. D. Tanksalvala, C. S. Bevis, V. H. Vartanian, H. C. Kapteyn, D. E. Adams, and M. M. Murnane, , Nano Lett. 16 5444 (2016).
7. R. Geneaux, H. J. B. Marroux, A. Guggenmos, D. M. Neumark and S. R. Leone, Philosophical Transactions of the Royal Society A 377 20170463 (2019).
8. H. Wang, M. Chini, S. Chen, C.-H. Zhang, F. He, Y. Cheng, Y. Wu, U. Thumm, and Z. Chang, Phys. Rev. Lett. 105 143002 (2010).
9. P. Peng, C. Marceau, M. Hervé, P.B. Corkum, A. Yu. Naumov and D. M. Villeneuve, Nat. Comm. 10 5269 (2019).
10. E. Goulielmakis, M. Schultze, M. Hofstetter, V. S. Yakovlev, J. Gagnon, M. Uiberacker, A. L. Aquila, E. M. Gullikson, D. T. Attwood, R. Kienberger, F. Krausz and U. Kleineberg, Science 320 1614 (2008).
11. E. A. Khazanov, Quant. Elec. 52 208 (2022).
12. M. Dorner-Kirchner, V. Shumakova, G. Coccia, E. Kaksis, B. E. Schmidt, V. Pervak, A. Pugzlys, A. Baltuška, M. Kitzler-Zeiler and P. A. Carpeggiani, ACS Photonics 10 84 (2023).
13. P. B. Corkum, N. H. Burnett and M. Y. Ivanov, Opt. Lett. 19 1870 (1994).
14. V. T. Platonenko and V. V. Strelkov, J. Opt. Soc. Am. B 16 435 (1999).
15. O. Tcherbakoff, E. Mével, D. Descamps, J. Plumridge and E. Constant, Phys. Rev. A 68 043804 (2003).
16. B. Shan, S. Ghimire and Z. Chang, J. Mod. Opt. 52 277 (2005).
17. Z. Chang, Phys. Rev. A 76 051403(R) (2007).
18. X. Feng, S. Gilbertson, H. Mashiko, H. Wang, S. D. Khan, M. Chini, Y. Wu, K. Zhao, and Z. Chang, Phys. Rev. Lett. 103 183901 (2009).
19. A. Fleischer and N. Moiseyev, Phys. Rev. A 74 053806 (2006).
20. H. Merdji, T. Auguste, W. Boutu, J.-P. Caumes, B. Carré, T. Pfeifer, A. Jullien, D. M. Neumark and S. R. Leone, Opt. Lett. 32 3134 (2007).
21. A. Fleischer, O. Kfir, T. Diskin, P. Sidorenko and O. Cohen, Nat. Photon. 8 543 (2014).
22. A. Fleischer, E. Bordo, O. Kfir, P. Sidorenko and O. Cohen, J. Phys. B 50 034001 (2017).
23. E. Ragonis, E. Ben-Arosh, L. Merensky and A. Fleischer, in preparation.
24. M. Nisoli, S. DeSilvestri and O. Svelto, Appl. Phys. Lett. **68**, 2793 (1996).
25. J. B. Watson, A. Sanpera and K. Burnett, Phys. Rev. A 51 1458 (1995).
26. H. J. Shin, D. G. Lee, Y. H. Cha, K. H. Hong and C. H. Nam, Phys. Rev. Lett. 83 2544 (1999).
27. E. Brunetti, R. Issac, and D. A. Jaroszynski, Phys. Rev. A 77 023422 (2008).
28. X.-B. Bian and A. D. Bandrauk, Appl. Sci. 3 267 (2013).
29. A. Dubrouil, O. Hort, F. Catoire, D. Descamps, S. Petit1, E. Mével, V.V. Strelkov and E. Constant, Nat. Comm. 5 4637 (2014).
30. C. M. Heyl, J. Güdde, U. Höfer, and A. L'Huillier, Phys. Rev. Lett. 107 033903 (2011).
31. X. He, M. Miranda, J. Schwenke, O. Guilbaud, T. Ruchon, C. Heyl, E. Georgadiou, R. Rakowski, A. Persson, M. B. Gaarde, and A. L'Huillier, Phys. Rev. A 79 063829 (2009).
32. A. Zaïr, M. Holler, A. Guandalini, F. Schapper, J. Biegert, L. Gallmann, U. Keller, A. Wyatt, A. Monmayrant, I. A. Walmsley, E. Cormier, T. Auguste, J. P. Caumes and P. Salières, Phys. Rev. Lett. 100 143902 (2008).

## References (including paper titles)

1. P. Corkum and F. Krausz, "Attosecond Science", Nat. Phys. **3** 381 (2007).
2. F. Krausz and M. Ivanov, "Attosecond Physics", Rev. Mod. Phys. **81** 545 (2009).
3. J. Li, J. Lu, A. Chew, S. Han, J. Li, Y. Wu, H. Wang, S. Ghimire and Z. Chang, "Attosecond science based on high harmonic generation from gases and solids", Nat. Comm. 11 2748 (2020).
4. O. Kfir, S. Zayko, C. Nolte, M. Sivis, M. Möller, B. Hebler, S. S. P. K. Arekapudi, D. Steil, S. Schäfer, M. Albrecht, O. Cohen, S. Mathias and Claus Ropers, "Nanoscale magnetic imaging using circularly polarized high-harmonic radiation", Science advances 3 eaao4641 (2017).
5. J. Miao, T. Ishikawa, I. K. Robinson, M. M. Murnane, "Beyond crystallography: Diffractive imaging using coherent X-ray light sources", Science 348, 530 (2015).
6. E. R. Shanblatt, C. L. Porter, D. F. Gardner, G. F. Mancini, R. M. Karl, Jr., M. D. Tanksalvala, C. S. Bevis, V. H. Vartanian, H. C. Kapteyn, D. E. Adams, and M. M. Murnane, "Quantitative Chemically Specific Coherent Diffractive Imaging of Reactions at Buried Interfaces with Few Nanometer Precision", Nano Lett. 16 5444 (2016).
7. R. Geneaux, H. J. B. Marroux, A. Guggenmos, D. M. Neumark and S. R. Leone, "Transient absorption spectroscopy using high harmonic generation: a review of ultrafast X-ray dynamics in molecules and solids", Philosophical Transactions of the Royal Society A 377 20170463 (2019).
8. H. Wang, M. Chini, S. Chen, C.-H. Zhang, F. He, Y. Cheng, Y. Wu, U. Thumm, and Z. Chang, "Attosecond Time-Resolved Autoionization of Argon", Phys. Rev. Lett. 105 143002 (2010).
9. P. Peng, C. Marceau, M. Hervé, P.B. Corkum, A. Yu. Naumov and D. M. Villeneuve, "Symmetry of molecular Rydberg states revealed by XUV transient absorption spectroscopy", Nat. Comm. 10 5269 (2019).
10. E. Goulielmakis, M. Schultze, M. Hofstetter, V. S. Yakovlev, J. Gagnon, M. Uiberacker, A. L. Aquila, E. M. Gullikson, D. T. Attwood, R. Kienberger, F. Krausz and U. Kleineberg, "Single Cycle Nonlinear Optics", Science 320 1614 (2008).
11. E. A. Khazanov, "Post-compression of femtosecond laser pulses using self-phase modulation: from kilowatts to petawatts in 40 years", Quant. Elec. 52 208 (2022).
12. M. Dorner-Kirchner, V. Shumakova, G. Coccia, E. Kaksis, B. E. Schmidt, V. Pervak, A. Pugzlys, A. Baltuška, M. Kitzler-Zeiler and P. A. Carpeggiani, "HHG at the Carbon K-Edge Directly Driven by SRS Red-Shifted Pulses from an Ytterbium Amplifier", ACS Photonics 10 84 (2023).
13. P. B. Corkum, N. H. Burnett and M. Y. Ivanov, "Subfemtosecond pulses", Opt. Lett. 19 1870 (1994);
14. V. T. Platonenko and V. V. Strelkov, "Single attosecond soft-x-ray pulse generated with a limited laser beam, J. Opt. Soc. Am. B 16 435 (1999).
15. O. Tcherbakoff, E. Mével, D. Descamps, J. Plumridge and E. Constant, "Time-gated high-order harmonic generation", Phys. Rev. A 68 043804 (2003).
16. B. Shan, S. Ghimire and Z. Chang, "Generation of the attosecond extreme ultraviolet supercontinuum by a polarization gating", J. Mod. Opt. 52 277 (2005).
17. Z. Chang, "Controlling attosecond pulse generation with a double optical gating", Phys. Rev. A 76 051403 (R) (2007).
18. X. Feng, S. Gilbertson, H. Mashiko, H. Wang, S. D. Khan, M. Chini, Y. Wu, K. Zhao, and Z. Chang, "Generation of Isolated Attosecond Pulses with 20 to 28 Femtosecond Lasers", Phys. Rev. Lett. 103 183901 (2009).
19. A. Fleischer and N. Moiseyev, "Attosecond laser pulse synthesis using bichromatic high-order harmonic generation", Phys. Rev. A 74 053806 (2006).
20. H. Merdji, T. Auguste, W. Boutu, J.-P. Caumes, B. Carré, T. Pfeifer, A. Jullien, D. M. Neumark and S. R. Leone, "Isolated attosecond pulses using a detuned second-harmonic field", Opt. Lett. 32 3134 (2007).
21. A. Fleischer, O. Kfir, T. Diskin, P. Sidorenko and O. Cohen, "Spin angular momentum and tunable polarization in high harmonic generation", Nat. Photon. 8 543 (2014).
22. A. Fleischer, E. Bordo, O. Kfir, P. Sidorenko and O. Cohen, "Polarization-fan high-order harmonics", J. Phys. B 50 034001 (2017).
23. E. Ragonis, E. Ben-Arosh, L. Merensky and A. Fleischer, in preparation.
24. M. Nisoli, S. DeSilvestri and O. Svelto, "Generation of high energy 10 fs pulses by a new pulse compression technique", Appl. Phys. Lett. **68**, 2793 (1996).
25. J. B. Watson, A. Sanpera and K. Burnett, "Pulse-shape effects and blueshifting in the single-atom harmonic generation from neutral species and ions", Phys. Rev. A 51 1458 (1995).
26. H. J. Shin, D. G. Lee, Y. H. Cha, K. H. Hong and C. H. Nam, "Generation of Nonadiabatic Blueshift of High Harmonics in an Intense Femtosecond Laser Field", Phys. Rev. Lett. 83 2544 (1999).
27. E. Brunetti, R. Issac, and D. A. Jaroszynski, "Quantum path contribution to high-order harmonic spectra", Phys. Rev. A 77 023422 (2008).
28. X.-B. Bian and A. D. Bandrauk, "Spectral Shifts of Nonadiabatic High-Order Harmonic Generation", Appl. Sci. 3 267 (2013).
29. A. Dubrouil, O. Hort, F. Catoire, D. Descamps, S. Petit1, E. Mével, V.V. Strelkov and E. Constant, "Spatio–spectral structures in high-order harmonic beams generated with Terawatt 10-fs pulses", Nat. Comm. 5 4637 (2014).
30. C. M. Heyl, J. Güdde, U. Höfer, and A. L'Huillier, " Spectrally Resolved Maker Fringes in High-Order Harmonic Generation", Phys. Rev. Lett. 107 033903 (2011).
31. X. He, M. Miranda, J. Schwenke, O. Guilbaud, T. Ruchon, C. Heyl, E. Georgadiou, R. Rakowski, A. Persson, M. B. Gaarde, and A. L'Huillier, "Spatial and spectral properties of the high-order harmonic emission in argon for seeding applications", Phys. Rev. A 79 063829 (2009).
32. A. Zaïr, M. Holler, A. Guandalini, F. Schapper, J. Biegert, L. Gallmann, U. Keller, A. Wyatt, A. Monmayrant, I. A. Walmsley, E. Cormier, T. Auguste, J. P. Caumes and P. Salières, " Quantum Path Interferences in High-Order Harmonic Generation", Phys. Rev. Lett. 100 143902 (2008).